\documentclass{emulateapj}
\usepackage{psfrag}
\usepackage{epsfig}
\usepackage{natbib}
\usepackage{graphicx}
\newcommand{\eqb}{\begin{eqnarray}}
\newcommand{\eqe}{\end{eqnarray}}
\newcommand{\pdiff}[2]{\frac{\partial #1}{\partial #2}}
\newcommand{\pmax}{p_{\rm max}}
\newcommand{\dif}[2]{\frac{{\rm d} #1}{{\rm d} #2}}
\newcommand{\diff}{{\rm d}}

\shorttitle{Steady-state solutions in NL-DSA}
\shortauthors{Reville, Kirk and Duffy}

\begin{document}

\title{Steady-State Solutions in Nonlinear Diffusive Shock Acceleration}

\author{B. Reville and J.G. Kirk}
\affil{Max-Planck-Institut f\"ur Kernphysik, Heidelberg 69029, Germany}

\and

\author{P. Duffy}
\affil{UCD School of Physics, University College Dublin, Dublin 4, Ireland}

\begin{abstract}
  Stationary solutions to the equations of non-linear diffusive shock
  acceleration play a fundamental role in the theory of cosmic-ray
  acceleration. Their existence usually requires that a fraction of
  the accelerated particles be allowed to escape from the system.
  Because the scattering mean-free-path is thought to be an increasing
  function of energy, this condition is conventionally implemented as
  an upper cut-off in energy space --- particles are then permitted to
  escape from any part of the system, once their energy exceeds this
  limit. However, because accelerated particles are responsible for
  substantial amplification of the ambient magnetic field in a region upstream of
  the shock front, we examine an alternative approach in which
  particles escape over a spatial boundary. We use a
  simple iterative scheme that constructs stationary numerical
  solutions to the coupled kinetic and hydrodynamic equations.  For
  parameters appropriate for supernova remnants, we find stationary
  solutions with efficient acceleration when the escape boundary is
  placed at the point where growth and advection of 
  strongly driven non-resonant waves are
  in balance. We also present the energy dependence of the
  distribution function close to the energy where it cuts off - a
  diagnostic that is in principle accessible to observation.
\end{abstract}


\section{Introduction }
\label{Introduction}

It is widely thought that the diffusive acceleration of charged particles 
at shock fronts can be very efficient 
\citep[for a review see][]{MalkovDrury}. 
The precise value of the efficiency is controlled by the microphysics of the 
injection of low-energy particles into the acceleration process, i.e., by 
processes operating on small length-scales. However,
in the case of non-relativistic shocks, such 
as those encountered in supernova remnants, 
the bulk of the energy in nonthermal particles
is carried off by those of the highest attainable energy,
i.e., by particles 
that interact with relatively large length-scale structures. 
 
Conventionally, the physics of this system is captured by combining 
a hydrodynamic description of the background plasma with the 
diffusion-advection equation obeyed by the distribution function of the 
accelerated particles. 
Analysis of 
the {\em stationary} solutions of these equations is the foundation on 
which the study of the overall efficiency of the process and the maximum energy 
to which particles can be accelerated rests.  

The importance of stationary solutions was realized early on by 
\citet{DruryVolk81} who used a reduced, two-fluid description to analyze 
acceleration by plane shock fronts in a one-dimensional flow. 
Provided accelerated particles are not permitted to leave 
the system, the two-fluid description can be derived from the
full description, including the diffusion-advection equation, using only 
two plausible assumptions
about the particle distribution function. 
The more important of these is
that the so-called {\em effective diffusion coefficient} is always positive.
\citet{DruryVolk81} proved that this restricts the possible stationary 
flow patterns to those containing a {\em precursor} in which the accelerated 
particles decelerate the incoming upstream flow, followed by 
a hydrodynamic shock 
front. This important
result, which implies that only a finite number of stationary
solutions exist for shocks of a given Mach number, 
was subsequently generalized to the relativistic case by \citet{baringkirk91}. 

However, the relevance of these studies is questionable if
the diffusion coefficient is an increasing function of particle energy,
as is the case, for example, if the transport is dominated by 
scattering off Alfv\'en waves via the cyclotron resonance.
One reason for this is that 
the timescale on which a stationary solution is approached becomes 
large at high particle energy. In this case, quasi-stationary solutions 
with a constant distribution function below 
a slowly evolving upper cut-off in energy can be expected to establish
themselves,  
and have been found numerically by \citet{Bell87} and by
\citet{FalleGiddings}.
Another reason is that, 
as particles are accelerated to higher and higher energy,
their mean-free path increases, and at some point becomes comparable to 
the size of any realistic system. Thus, even if 
one considers only strictly stationary solutions, 
the escape of high-energy particles 
appears to be an important property that will limit the maximum energy to 
which particles can be accelerated in any given system.

One way of accounting for particle escape is to truncate the distribution
function above some finite value of the energy. 
But even with this simplification, finding stationary solutions of the 
combined kinetic and hydrodynamic
equations is much more difficult than solving the two-fluid system. 
An important advance
was the discovery of an approximate analytic solution of the
diffusion-advection equation by \cite{Malkov97}. This solution has a
particle distribution function that vanishes above an upper cut-off,
$\pmax$, in the magnitude of the particle momentum. Wherever the
hydrodynamic flow is compressed (usually throughout the precursor and
at the shock front) there exists a flux of particles across this
boundary in momentum space. These particles cease to contribute to the
stress-energy tensor, and, therefore, effectively escape from the
system. A similar distribution was adopted by \citet{achterberg87} 
when investigating numerical solutions to this equation. 
Under this assumption it is possible to find approximate analytic solutions, 
and relatively straightforward to construct numerical solutions to the
full set of equations \citep{achterberg87,MDV00,Blasi02,AB05}.

Two problems intrinsic to this approach
are immediately obvious. 
Firstly, 
the momentum dependence of the distribution close to $\pmax$ --- a 
diagnostic that is, in principle, accessible to observation --- is not 
well-approximated.
Secondly, the exclusion of a post-cursor, although made plausible by the 
two-fluid approach, is not always justified. 
The conditions that must
be fulfilled by the distribution function and the momentum-dependent 
diffusion coefficient for this assumption to be valid \citep{kirk90}
can in principle be checked {\em a posteriori}, but this is not 
straightforward for discontinuous distributions. 

These problems arise because an upper cut-off in momentum is used to
describe the physics of particle escape. Recently, however,
observational evidence has been accumulating suggesting that 
cosmic rays are responsible for substantial amplification of the
ambient magnetic field in the precursors of shock fronts in
supernova remnants \citep{Hwang,VinkLam,Bamba,Uchiyama}. 
This implies that the scattering
mean-free-path is not only a function of energy, but also depends
strongly on position with respect to the shock front. It, therefore, highlights 
a more serious short-coming of the approach that uses a
cut-off in momentum space: The amplification of the field is likely 
to be connected with
the spatial flux of energetic particles, which is artificially distorted 
if particles are assumed to vanish across a momentum boundary.
In this paper we examine 
stationary solutions with
escape through a spatial boundary instead of through a boundary in
energy space. 

Spatial boundaries have previously been implemented in Monte-Carlo 
simulations of the non-linear acceleration problem.
In particular, \citet{VEB06} used 
a model equation to describe Alfv\'enic
turbulence and coupled it to a Monte-Carlo simulation that permitted escape
over a spatial boundary. 
In this way they were able to investigate 
the effects of an enhanced resonant
interaction between the turbulent waves and the accelerated particles,
although the position of the boundary itself remained arbitrary.

Recently, 
\cite{ZirPtusk08} used MHD simulations to describe the excited turbulence and
find the diffusion coefficient of the highest energy
accelerated particles. They allowed these particles
to escape over a spatial boundary, but used the test-particle approximation
in which the flow speed is unaffected by the particles. 
In discussing shock fronts in supernova remnants, they suggested that 
the boundary leads the shock front by a distance
given roughly by the radius of the remnant.
 
Our approach differs from these, 
not only in the numerical method used to solve the 
acceleration problem, but
also in the input physics. 
Resonant interactions between energetic particles and Alfv\'en waves
were long thought to be responsible for coupling these particles to the 
background plasma \citep{bell78,voelkmckenzie,achterberg83,belllucek00}. 
However, well ahead of the shock front, non-resonant processes are 
more strongly driven and can be expected to dominate under the conditions
present in supernova remnants 
\citep{bell04,Pelletier,bykovtoptyghin,revilleetal07,ZPV08}.
In the linear phase, these instabilities 
inject short-wavelength turbulence into the 
plasma, resulting in a relatively large diffusion coefficient that is 
proportional to the square of the particle momentum
\citep{ZirPtusk08}. However, the non-linear evolution includes not 
only a cascade of energy to smaller length scales, but also the 
appearance of large-scale structures such as cavities \citep{bell05}. 
In this regime, the mean-free-path is reduced, and its $p^2$
dependence eliminated, as can be seen from 
large-scale numerical simulations 
of the transport properties \citep{revilleetal08}.
These simulations 
have not yet advanced to the stage where they can provide a model diffusion 
coefficient over a wide dynamic range of momentum.  
Consequently, we model
this effect by  
assuming the diffusion to be of Bohm type in a background field 
that is amplified to the value at which the non-resonant instabilities 
are expected to saturate. 
Although this is a relatively crude approach, it 
enables us to solve the non-linear problem of finding stationary solutions. 
We are then able to check the location of the spatial boundary, which
should be located where the growth rate of 
the non-resonant modes
is approximately equal to the speed of advance of the shock
divided by the distance from the shock front. Since the
escaping flux is dominated by the highest energy particles,
we do not expect
that changing the form of the diffusion coefficient will 
alter significantly our conclusions, provided it remains an
increasing function of momentum.

The paper is set out as follows: 
In
Section~\ref{basic} we set up the advection-diffusion equation and the
hydrodynamic equations governing the system of accelerated particles
and background plasma. The two ways of allowing for particle escape
are discussed in Section~\ref{escape}.
In Section~\ref{codedescription} we
describe the iteration and finite difference methods used to find
stationary solutions of the combined advection-diffusion and
hydrodynamic equations using the two different boundary conditions,
and in Section~\ref{comparison} we compare and discuss the results.
A discussion of the self-consistency of the position of the 
spatial boundary and a summary of our conclusions is presented in 
Section~\ref{conclusions}.

\section{Basic equations}
 \label{basic}

 We consider a gas subshock located at $x=0$ with a flow profile, in
 the subshock rest frame, given by
\begin{eqnarray*}
 U(x)=\left\lbrace
\begin{array}{cl}
u_2 & x>0 \\
u(x) & x\leq 0 
\end{array}
\right.
\end{eqnarray*}
with $u(x=0^-)=u_1$ and $u_2$ constant in the absence of a
post-cursor. The gas velocity far upstream is denoted by $u_0$. We
assume that, as a result of scattering centres frozen into the flow,
energetic particles, with speeds $v\gg u(x)$, undergo diffusion with a
momentum-dependent diffusion coefficient $\kappa(p)$. These particles
are also advected with the flow, adiabatically compressed and injected
from the thermal background.  The isotropic part of their phase space
density obeys the transport equation \citep{Skilling75I}
\begin{equation}
\label{transportEqn}
\pdiff{f}{t}+
U\pdiff{f}{x}-\pdiff{\;}{x}
\left(\kappa\pdiff{f}{x}\right)=
\frac{1}{3}\dif{U}{x}p\pdiff{f}{p}+Q_0(x,p),
\label{transporteqn}
\end{equation}
where $Q_0$ describes the injection of particles into the acceleration 
process. For mono-energetic injection at the gas subshock 
\eqb
Q_0(x,p) &=& \frac{\eta n_{{\rm g},1}u_1}{4\pi p_0^2}
\delta(p-p_0)\delta(x),
\nonumber
\eqe
where the number density of gas particles entering the shock front is $n_{{\rm g},1}$
and $\eta$ is the fraction of entering particles 
that take part in the acceleration process. In our notation, the 
particle mass and momentum are $m$ and $mc p$, 
so that $p$, $p_0$ etc.\ are dimensionless, and 
the phase-space distribution function
$f$ has the dimensions of an inverse volume. 

Integrating 
Eq.~(\ref{transportEqn}) first across the shock and then across 
the injection momentum, it follows that
\begin{equation}
\label{Qinj}
 f_0(p_0)=\frac{3u_1}{\Delta u}
\frac{\eta n_{{\rm g},1}}{4\pi p_0^3}
\end{equation}
with $\Delta u= u_1-u_2$.  
An important restriction on this approach is that the distribution
function at the injection momentum $p_0$ must be approximately
isotropic. This requires that the velocity of these particles should be
several times greater than that of the upstream plasma. 
Since we will be interested primarily in shocks in 
supernova remnants, where $u_0/c\lesssim 1/30$ we require 
$p_0\ge0.1$.

In this paper we are interested in steady state solutions to the
particle transport equation, and in particular the role played by
particle escape, when the pressure associated with the energetic
particles
\eqb
\label{Pcr}
 P_{\rm cr}(x)&=&\frac{4\pi}{3}mc^2\int_{p_0}^{\infty}
vp^3 f(x,p) \diff p
\eqe
reacts on the flow.
Sufficiently far upstream, in the absence of cosmic rays,
the gas has a density $\rho_0$ and pressure $P_{g,0}$. 
Mass and momentum conservation give
\begin{eqnarray}
\label{mass}
 \rho(x)u(x)&=&\rho_0u_0, \\
\label{momentum}
 P_{\rm cr}(x)+\rho(x) u(x)^2 +
 P_{g,0}\left({u(x)\over u_0}\right)^{-\gamma}
 &=&\rho_0 u_0^2+P_{g,0} ,
\end{eqnarray}
where $\gamma$ is the adiabatic index of the gas. 

The plasma flowing towards the shock is adiabatically compressed
and slowed in the precursor. It's velocity profile $u(x)$ 
is a monotonic function. 
Non-adiabatic heating is potentially
important \citep[e.g.][]{Caprioli08, VBE08}. 
However, in the interests of simplicity, we shall neglect it in the following.
For adiabatic heating alone, the sub-shock compression ratio
$r_{\rm s}=u_1/u_2$ and the pre-compression 
\eqb
R&=&u_0/u_1.
\label{precompression}
\eqe
are related by \citep[see, for example][]{LandauLifshitz}
\begin{equation}
r_{\rm s} = \frac{\gamma+1}{\gamma-1+2R^{\gamma+1}M^{-2}},
\end{equation}
where $M$ is the Mach number of the flow at upstream infinity.
For a given pre-compression ratio, we can use Bernoulli's equation
(\ref{momentum})
to determine the cosmic-ray pressure
at the shock
\begin{equation}
\label{Pshock}
{P_{\rm cr}(0^-)\over \rho_0u_0^2}=1-R^{-1}+\frac{1}{\gamma M^2}\left(1-R^{\gamma}\right).
\end{equation}

Thus, given the upstream conditions and $R$, or, alternatively, $r_{\rm s}$,  
we can determine the cosmic-ray pressure
at the shock.

\section{Particle escape}
\label{escape}

\subsection{Boundary in energy}

As mentioned in Section \ref{Introduction}, 
most previous analytic and semi-analytic calculations 
adopted a precursor of infinite spatial extent.
Escape is permitted by assuming that 
particles with energy above a certain threshold decouple from 
the plasma. In essence, this is equivalent to assuming that
the mean-free-path to scattering is energy dependent, and, 
above a certain threshold, 
becomes large compared to the size of the system. 
In the resonant scattering scenario, this implies that
the relevant wave spectrum cuts-off above 
a certain wavelength.

Since the most energetic particles are highly relativistic
an upper cut-off in the energy is equivalent to one in 
the magnitude of the momentum.
For $p>p_{\rm max}$, particles escape from the system, and the 
approximate steady-state solution for $p<p_{\rm max}$ is 
\begin{equation}
\label{ftotal}
 f(x,p)=f_0(p)\exp\left[
\frac{q}{3\kappa}\left(1-\frac{u_1}{u_2}\right)
\int_x^0\diff x' u(x')\right],
\end{equation}
where $q=\partial \ln f_0 / \partial\ln p$. 
The additional factor of $(1-u_2/u_1)$ was
included in \cite{BACaprioli07} to match the boundary condition
for weakly modified shocks. The distribution at the shock, for monoenergetic 
injection, is 
\begin{equation}
\label{fshock}
 f_0(p)=\frac{3Rr_{\rm s}}{Rr_{\rm s}U(p)-1}
\frac{\eta n_0}{4\pi p_0^3}
\exp\left[
-\int_{p_0}^p \frac{\diff p'}{p'}
\frac{3Rr_{\rm s}U(p')}{Rr_{\rm s}U(p')-1}
\right]
\end{equation}
where $U(p)=u_p(p)/u_0$ with 
\begin{equation}
\label{up}
 u_p(p)=u_1-\frac{1}{f_0}
\int_{-\infty}^{0^-}\diff x\dif{u}{x}f(x,p).
\end{equation}
Following \cite{Blasi02} we take $U(p_0)\approx R^{-1}$ to give 
\begin{equation}
 \eta=  4\pi p_0^3 f_0(p_0)\frac{r_{s}-1}{3n_0Rr_{s}}
\end{equation}
in agreement with equation~(\ref{Qinj}).
In the results of this work, we
will adopt the dimensionless injection parameter, 
$\nu$, as defined in \cite{MDV00}, 
\begin{equation}
\nu={4\pi\over 3}{m c^2\over \rho_0u_0^2}p_0^4f_0(p_0),
\end{equation}
where again, $p$ is dimensionless.

\subsection{Boundary in space}

A free-escape boundary upstream implies that all particles 
that cross a surface placed at a distance $L_{\rm esc}$ 
upstream of the shock leave the system. 
Essentially, the particle mean-free-path and, therefore, the 
diffusion coefficient,
become infinite at the escape boundary due to the absence of
scattering waves beyond that point. 
Since we employ a diffusion approximation, this can be implemented 
by setting the isotropic part of the distribution to zero on the boundary. 
Whilst inside, we assume particles undergo Bohm diffusion,
\begin{equation}
\begin{array}{lr}
 \kappa = \kappa_0 p,  & x>-L_{\rm esc}
\end{array}
\end{equation}
Higher energy particles 
have a larger mean-free-path and, therefore, 
propagate from the shock to the boundary more easily than low 
energy ones. Since they then escape, a turn-down in the spectrum 
results. 
However, this is a smooth 
decrease, rather than an abrupt cut-off. It occurs close to a
momentum $p^*$ defined by
\begin{equation}
L_{\rm esc} = \kappa(p^*)/u_0,
\end{equation}

In the test particle limit, where the upstream flow is unmodified and 
constant, a straightforward calculation shows that the spectrum at 
the shock, $f_0(p)$, has a slope given by
\begin{equation}
\label{TPsol}
 -\dif{\ln f_0}{\ln p} = 
\frac{3 u_1}{\Delta u}
\frac{1}{[1-\exp(-L_{\rm esc}u_0/\kappa)]}
\end{equation}
At low momenta, $p \ll p^*$, the spectrum agrees with the 
standard test-particle result without escape upstream, 
but at high momenta $p\sim p^*$
the spectrum cuts off exponentially. It should be noted that some particles
with momenta $p \leq p^*$ also reach the escape boundary, and 
the spectrum will start to turn over at these values. 
Although the particle distribution function $f$ vanishes at the 
free-escape boundary, the particle flux remains finite, being proportional 
to $p^2\kappa\partial f/\partial x$. 
In the test-particle approximation it can be 
written, making the substitution $s=p/p^*$,
\eqb
-s^2\kappa{\partial f\over\partial x}
&=&\frac{u_0 f_0(p_0)s^2}{{\rm e}^{1/s}-1} \exp
\left[\frac{-3 r_{\rm s}}{r_{\rm s}-1}\int_{s_0}^s 
\frac{\diff \ln s'}{1-{\rm e}^{-1/s'}}\right]
\label{escapingflux}
\eqe  
where $s_0=p_0/p^*$. As a function of momentum, 
the escaping flux is sharply peaked at 
$s\approx (r_{\rm s}-1)/(r_{\rm s}+2)$.

\section{Method of solution}
\label{codedescription}

In the case of a boundary in energy, we use an iteration
scheme similar to that employed by \citet{AB05}.
Given a shock Mach number $M$, shock speed $u_0$ and momentum range
$p_{0}$, $p_{\rm max}$, we look for converged solutions for each value
of $R$. 
The initial flow profile is linear, and the spectrum is a single power
law. This allows us to calculate $U(p)$ 
from which we determine the injection parameter $\eta$ in 
 Eq~(\ref{fshock}), by identifying the cosmic ray pressure at the shock
front in Eq~(\ref{momentum}) with the integral over $p$ of $vp^3f_0$, as 
indicated by Eq~(\ref{Pcr}).
Using equations (\ref{ftotal}) and (\ref{Pcr}) we then update the flow
using the fluid equations. This gives new values for $u(x)$ and $q(p)$, and,
hence a new value for $\eta$. This is repeated until $\eta$ has converged.

Numerical solutions for steady state 
modified shocks and particle spectra, with a 
free escape boundary upstream, are also found using an iterative procedure. 
For a given Mach number ($M$), shock speed ($u_0$) and pre-compression 
ratio ($R$), we initialize the spectrum using the test particle 
solution. The flow profile is then adjusted using the flux 
conservation equations (\ref{mass},\ref{momentum}). With 
the modified flow, the distribution function is updated by solving the time 
dependent transport equation for particles,
equation (\ref{transportEqn}),
with the upstream boundary 
condition $f(-L_{\rm esc},p)=0$. 
For this we use a 
Crank-Nicholson scheme centred in space and upwind in momentum. 
Using Eq.~(\ref{Pshock}), the distribution function can
be normalized to match the pressure at the shock. When, at each iterative 
step, the transport equation is solved, the resulting cosmic ray pressure 
is used to update the flow which, in turn, is then used to update the 
particle distribution. 

The solution is found when the fluid quantities
no longer change and the injection parameter 
$\eta$ has converged. The code steps through 
values of the pre-compression ratios, using the
previous converged profile and spectrum as the initial 
conditions for the next iteration. In this manner we obtain a full 
set of solutions that depend on $\eta$, $M$, $R$ and $L_{\rm esc}$.

\section{Results}
\label{comparison}

\subsection{Comparison of approaches}

In order to make a meaningful comparison between the two methods we take
\begin{equation}
p^*=p_{\rm max}
\end{equation}
so that the boundary for upstream spatial escape, characterized by
$L_{\rm esc}$, is placed one diffusion length away from the shock for
particles with momentum $p_{\rm max}$.  Consequently, the gradual
momentum cut-off produced by the spatial boundary technique is
close to the position of the abrupt momentum boundary.

\begin{figure}
\begin{center}
\vbox{
   \psfrag{p1}[Bl][Bl][1.0][+0]{\hspace*{-0cm} {  $M=10$}}
   \psfrag{p2}[Bl][Bl][1.0][-0]{\hspace*{-0cm} {  $50$}}
   \psfrag{p3}[Bl][Bl][1.0][-0]{\hspace*{-0cm} {  $100$}}
   \psfrag{p4}[Bl][Bl][1.0][-0]{\hspace*{-0cm} {  $500$}}
   \psfrag{p5}[Bl][Bl][1.0][-0]{\hspace*{-0cm} { $R$}}
   \psfrag{p6}[Bl][Bl][1.0][-0]{\hspace*{-0cm} { $\nu$}}
   
\includegraphics[width=0.5\textwidth]{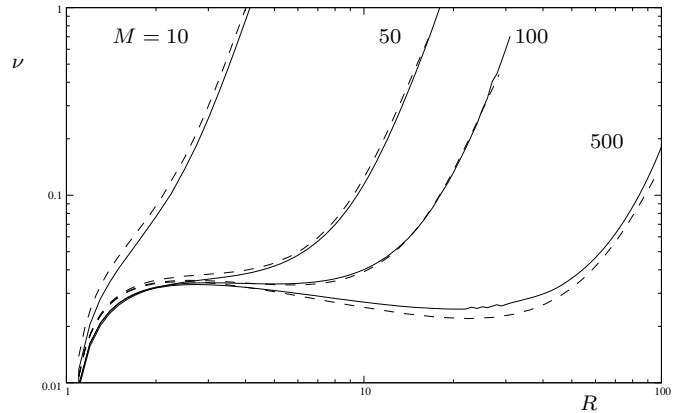}
}
\caption{
The injection parameter $\nu$ as a function of the 
pre-compression $R$ --- see Eq.~(\ref{precompression}) ---
for increasing Mach number at a fixed upstream
velocity $u_0=5,000$ km s$^{-1}$. In generating this plot, 
we set $p_0=0.1mc$ and $p^*=p_{\rm max}=10^3 mc$. 
The solid
lines are found using a spatial boundary, the dashed
lines use a boundary in energy.}
\label{fig1}
\end{center}
\end{figure}

\begin{figure}
\begin{center}
\vbox{
   \psfrag{p1}[Bl][Bl][1.0][+0]{\hspace*{-0cm} { $\nu$}}
   \psfrag{p2}[Bl][Bl][1.0][-0]{\hspace*{-0cm} { $R$}}
   
\includegraphics[width=0.5\textwidth]{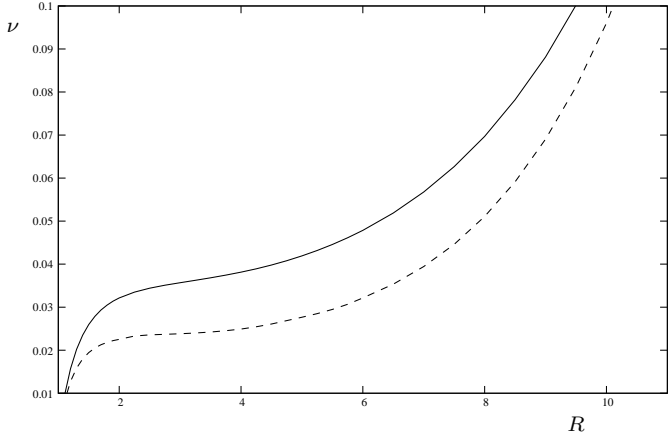}
}
\caption{
The injection parameter $\nu$ as a function of the pre-compression ratio $R$
for a $M=50$ shock 
using different injection and escape momenta..
Solid: $p_0=0.1mc$, $p^*=10^3mc$, dashed:
$p_0=0.1mc$, $p^*=10^4mc$.
}
\label{fig2}
\end{center}
\end{figure}

Figure \ref{fig1} plots the numerically determined injection
parameter $\nu$ as a function of the pre-compression ratio for the
two different methods. These plots follow a similar form to those
studied analytically by \cite{Malkovb97}. The curve for each shock
Mach number can, in general, be separated into three different
regimes. The inefficient weakly modified regime where the spectrum
resembles the test particle solution, the efficient regime where the
shock is strongly modified, with a weak subshock, and an intermediate
regime between the two. This intermediate regime lies typically in a
narrow range of the injection parameter $\nu$.  As the shock Mach number
increases, multiple solutions for a single value of $\nu$ 
occur,
as found by \citet{Malkovb97,ABG08}.  The two approaches give 
similar results, as expected, 
since the normalization of the spectrum is fixed by
Eq~(\ref{Pshock}). The differences arise due to the shape of the
spectrum close to the cut-off. In Fig~\ref{fig2} 
we show the $\nu-R$ diagram for a Mach $50$ shock for
different values of maximum and minimum momentum using the free-escape
boundary.

\begin{figure*}
\centering
\vbox{
\hbox{
   \psfrag{px}[Bl][Bl][1.0][-0]{\hspace*{-0cm} { $-x$}}
   \psfrag{py}[Bl][Bl][1.0][-0]{\hspace*{-0cm} {  $P_{cr}$}}
   \psfrag{u1}[Bl][Bl][1.0][-0]{\hspace*{-0cm} { $-x$}}
   \psfrag{u2}[Bl][Bl][1.0][-0]{\hspace*{-0cm} {  $U$}}	
\includegraphics[width=0.45\textwidth]{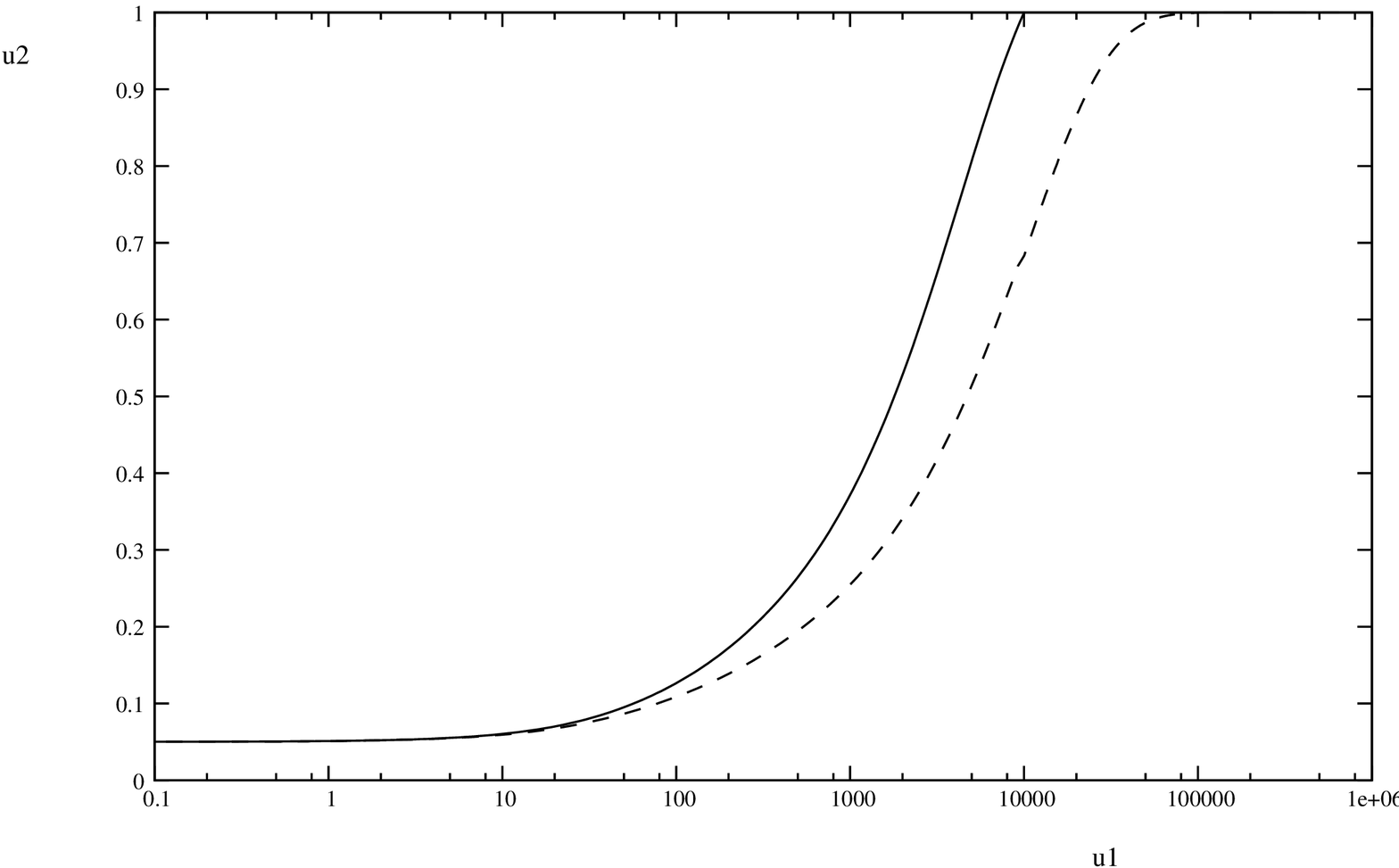}
\includegraphics[width=0.45\textwidth]{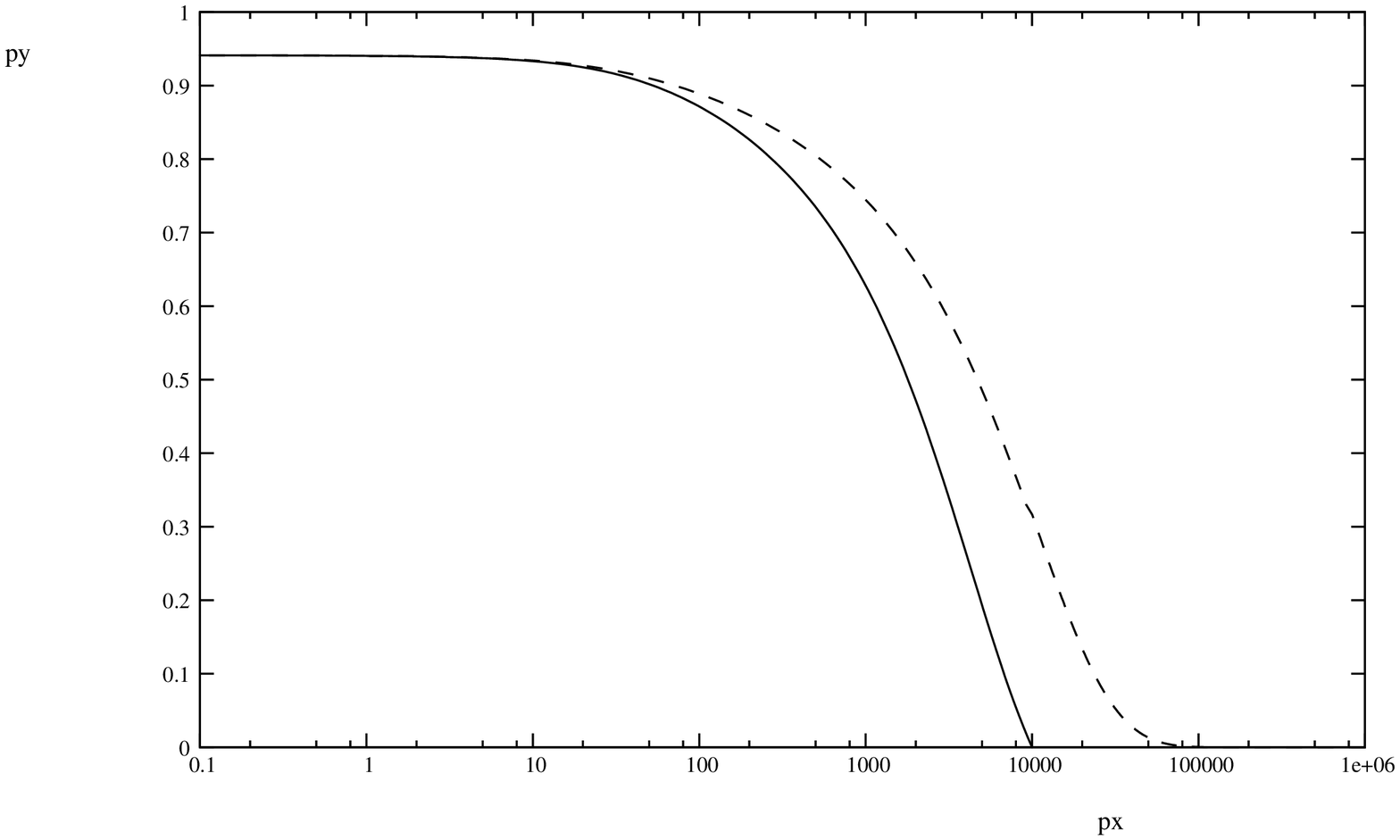}
}
\hbox{
   \psfrag{p1}[Bl][Bl][1.0][-0]{\hspace*{-0cm} { $p$}}
   \psfrag{p2}[Bl][Bl][1.0][-0]{\hspace*{-0cm} {  $p^4f$}}
   \psfrag{q1}[Bl][Bl][1.0][-0]{\hspace*{-0cm} { $p$}}
   \psfrag{q2}[Bl][Bl][1.0][-0]{\hspace*{-0cm} { $q$}}
\includegraphics[width=0.45\textwidth]{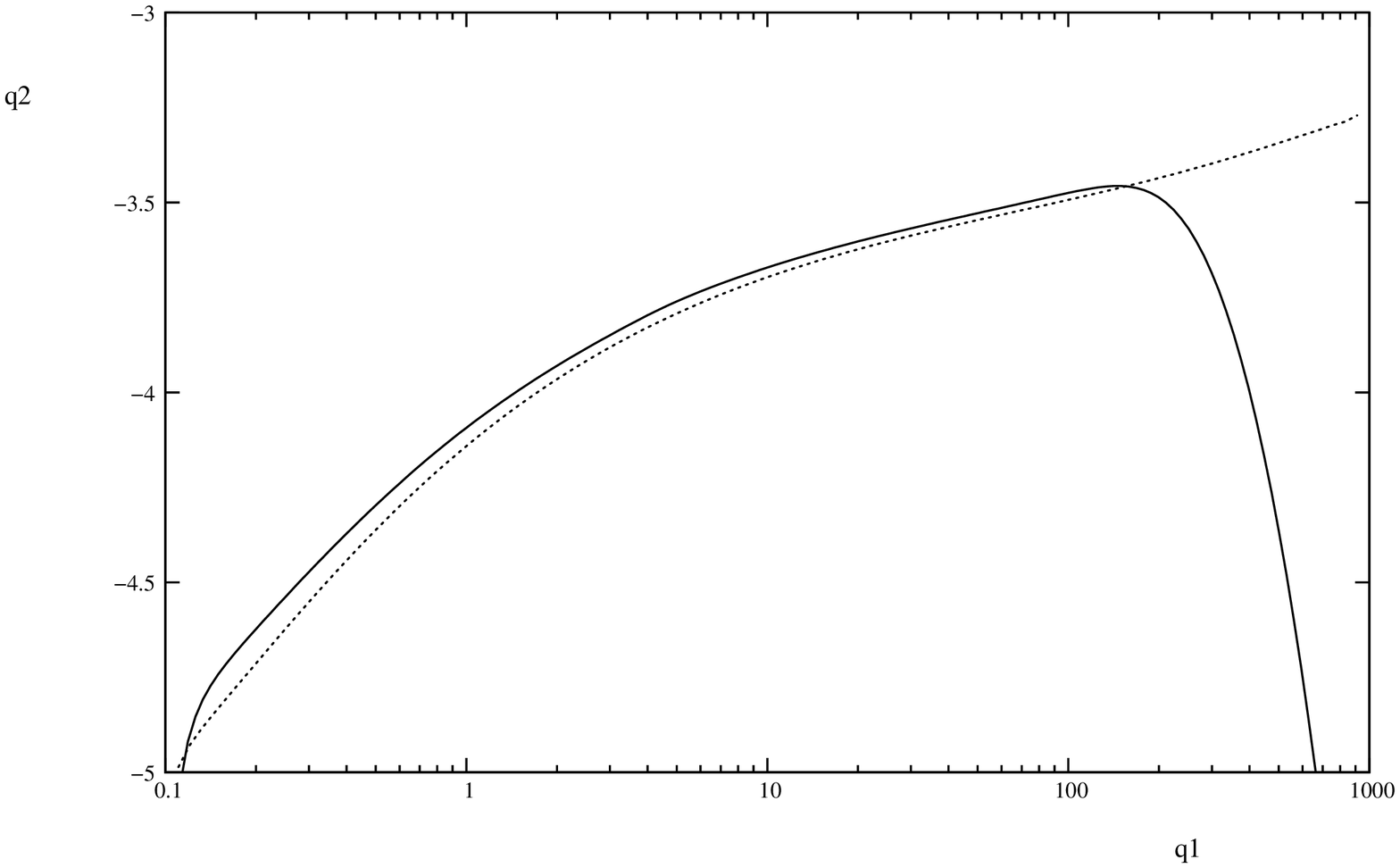}	
\includegraphics[width=0.45\textwidth]{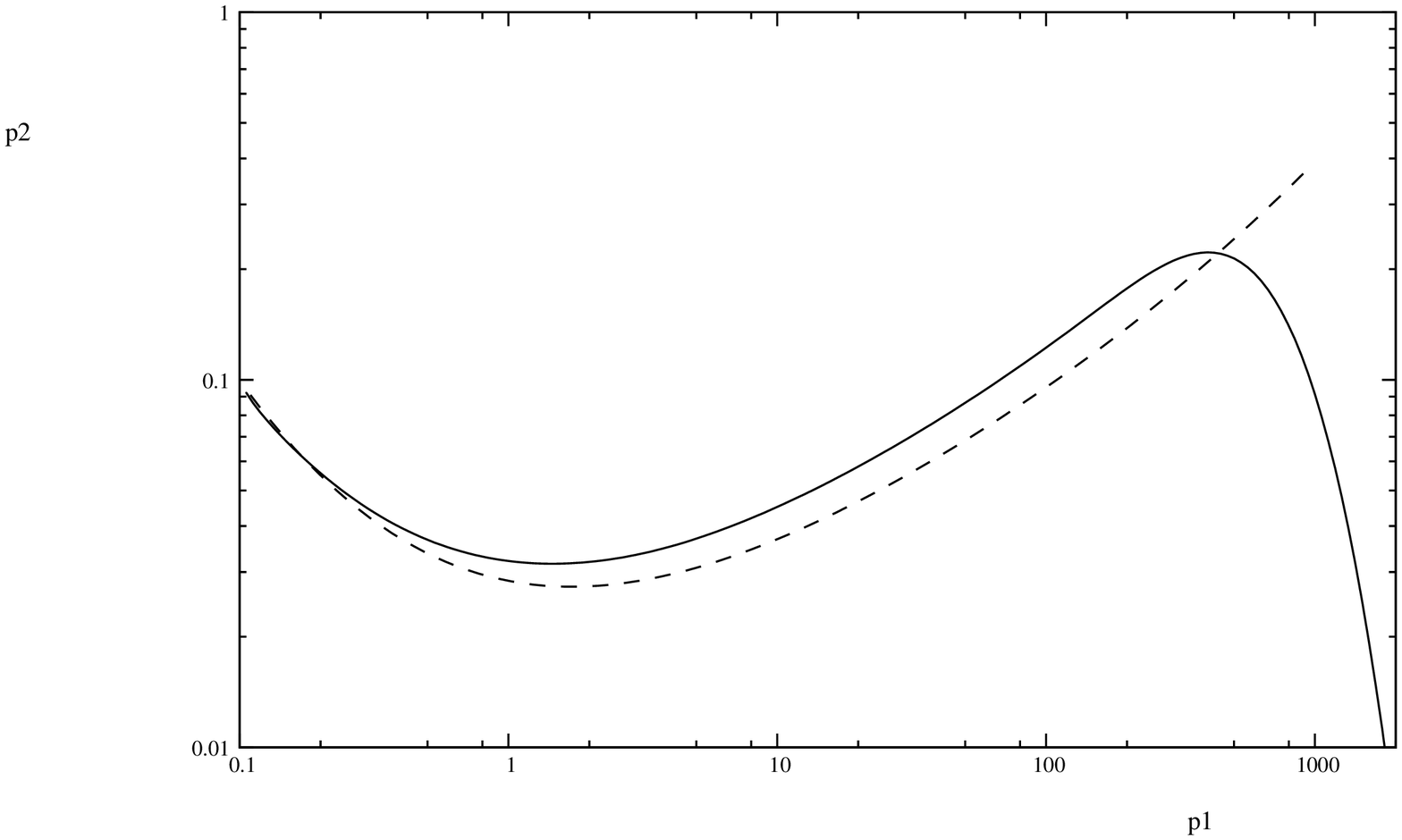}

}

}
\caption{
Normalized cosmic-ray pressure and flow profile
for a $M=100$ shock, as a function of distance upstream of the
sub shock. The pre-cursor compression ratio is $R=20$.
The solid lines correspond to the spatial boundary approach
and dashed lines to the momentum boundary.
}
\label{fig3}
\end{figure*}

\begin{figure*}
\centering
\vbox{
\hbox{
   \psfrag{px}[Bl][Bl][1.0][-0]{\hspace*{-0cm} { $-x$}}
   \psfrag{py}[Bl][Bl][1.0][-0]{\hspace*{-0cm} {  $P_{cr}$}}
   \psfrag{u1}[Bl][Bl][1.0][-0]{\hspace*{-0cm} {$-x$}}
   \psfrag{u2}[Bl][Bl][1.0][-0]{\hspace*{-0cm} {  $U$}}	
\includegraphics[width=0.45\textwidth]{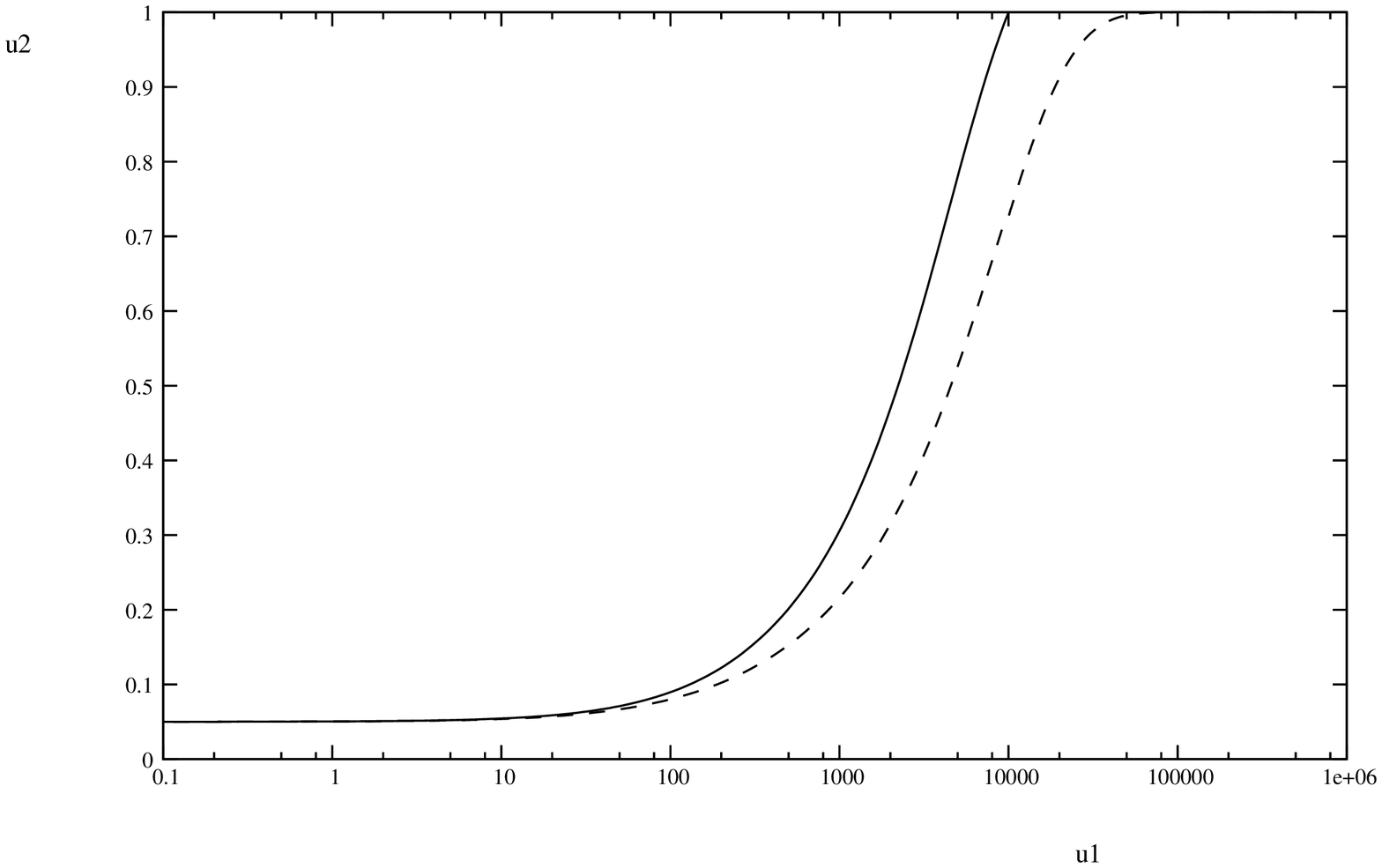}
\includegraphics[width=0.45\textwidth]{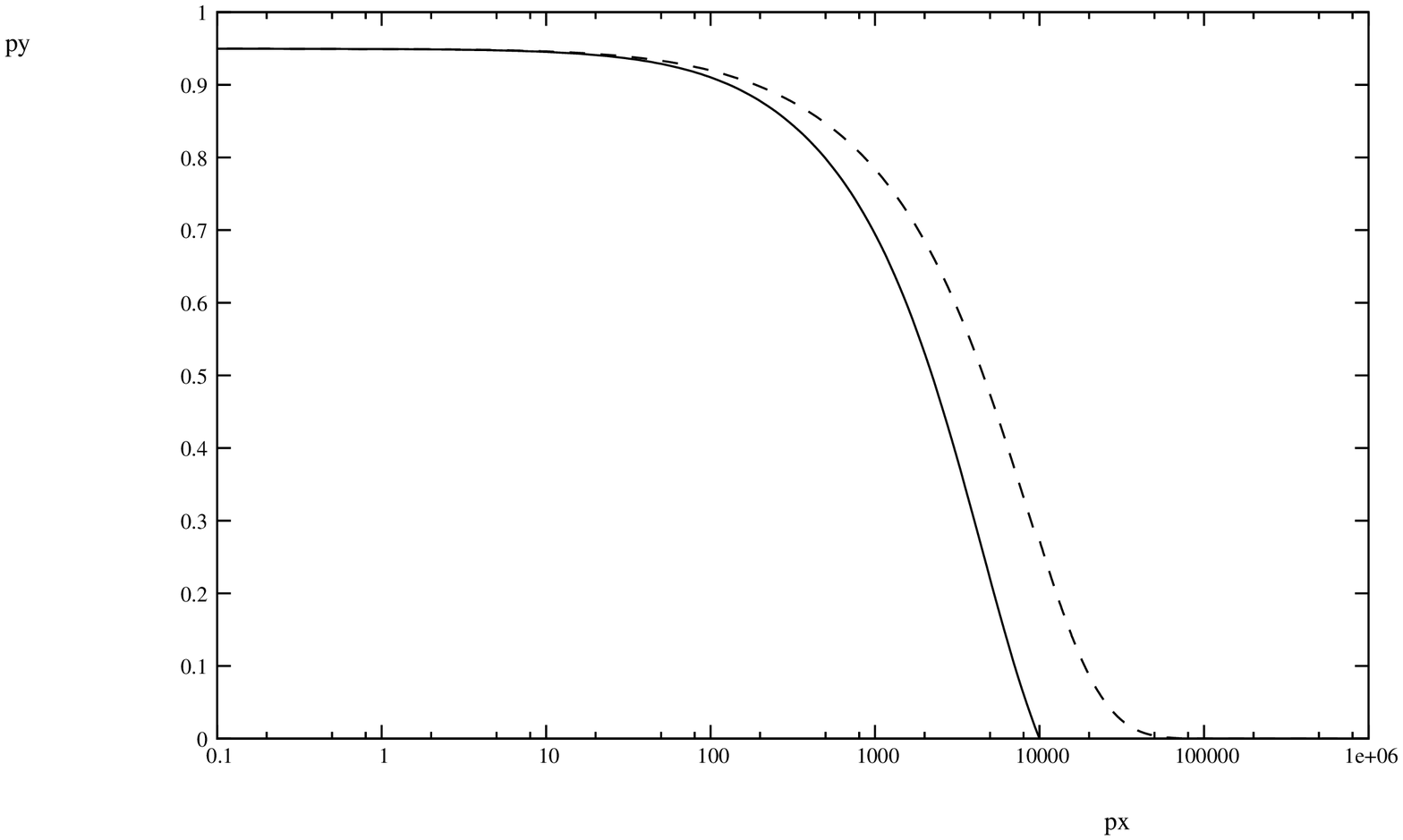}
}
\hbox{
   \psfrag{p1}[Bl][Bl][1.0][-0]{\hspace*{-0cm} { $p$}}
   \psfrag{p2}[Bl][Bl][1.0][-0]{\hspace*{-0cm} {  $p^4f$}}
   \psfrag{q1}[Bl][Bl][1.0][-0]{\hspace*{-0cm} {$p$}}
   \psfrag{q2}[Bl][Bl][1.0][-0]{\hspace*{-0cm} {$q$}}	
\includegraphics[width=0.45\textwidth]{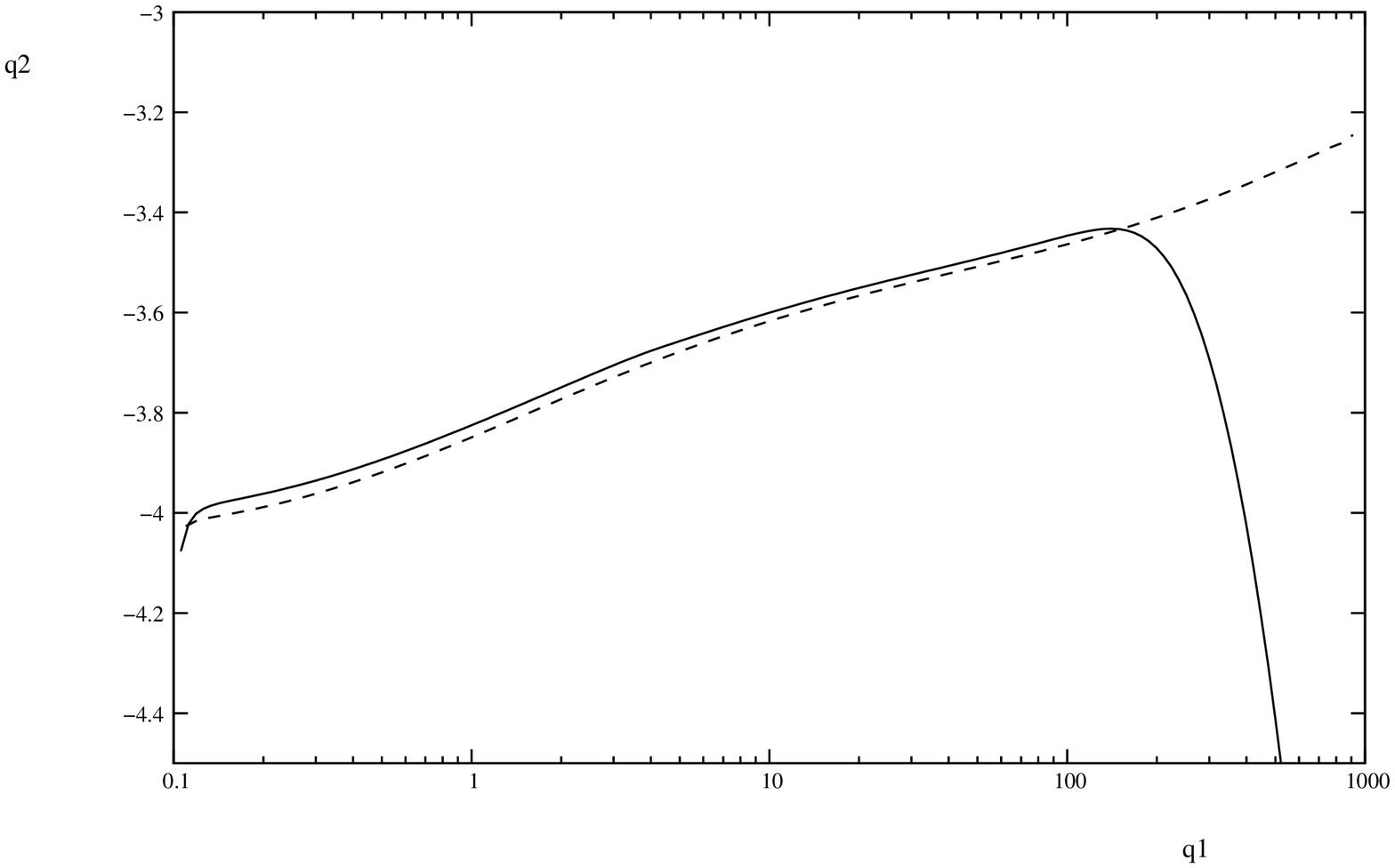}
\includegraphics[width=0.45\textwidth]{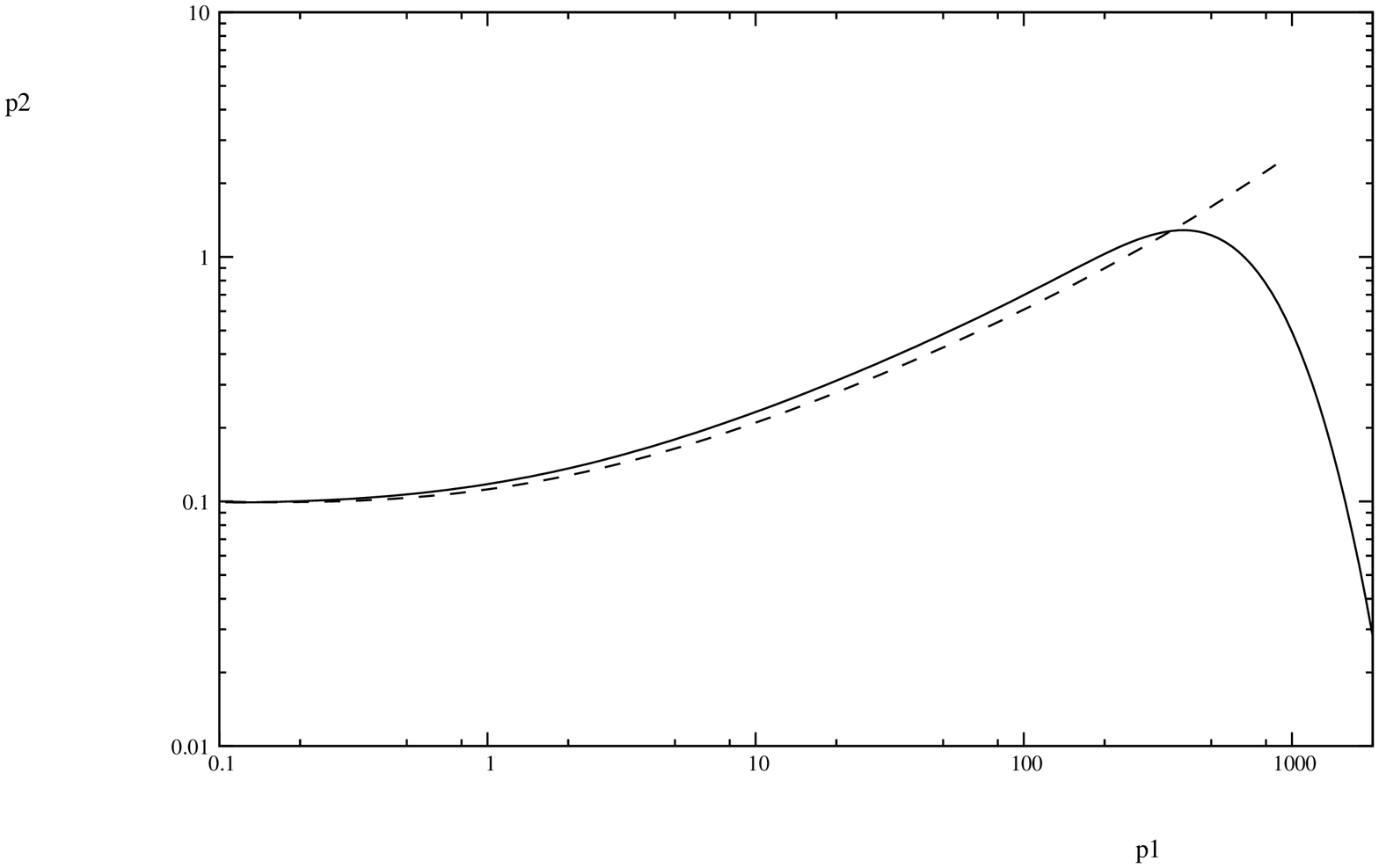}

}

}
\caption{
Normalized cosmic-ray pressure and flow profile
for a $M=500$ shock, as a function of distance upstream of the
sub shock. The pre-cursor compression ratio is $R=20$.
The solid lines correspond to the spatial boundary approach
and dashed lines to the momentum boundary.
}
\label{fig4}
\end{figure*}

Figures \ref{fig3} and \ref{fig4} compare the flow and pressure profiles,
as well as the spectrum and spectral index for Mach
numbers, $M=100$ and $M=500$. Both examples are with
a pre-compression ratio of
$R=20$. As can be seen from Fig.~\ref{fig1}, the two cases fall in
different regimes. For the $M=100$ case, the shock is strongly
modified with $r_s\approx2.1$, while the $M=500$ case is in the
intermediate regime with a subshock compression ratio of
$r_s\approx3.86$. This can be seen from the low energy shape
of the spectrum: the reduced sub-shock compression leads to a much softer
spectrum at the $M=100$ shock.

The flow profiles of the two models differ only slightly. This follows
naturally from the boundary conditions: with a free escape
boundary, the precursor length is fixed by $L_{\rm esc}$.
Using a boundary in momentum leads to a slightly extended
precursor, due to the larger number of particles with momenta 
$p^*$ interacting on a slightly larger scale.
For both Mach numbers, there is a smooth
turnover in the spectrum at momenta close to $p^*$. 
Furthermore, the momentum derivative of the distribution remains
negative in this range. In this case, it can be shown 
\citep[][Eq~(12)]{kirk90} that the dynamics of particles close to 
$p^*$ do not permit a post-cursor. However, the
possibility remains that the dynamics of the injection
mechanism may still do so \citep{zankwebbdon93}.
Particles with
momenta slightly less than $p^*$ are also able to diffuse a distance
$L_{\rm esc}$ from the shock and escape upstream. This is what leads
to the observed turn-over.   
The reduced value for $p^*$ is due to the
non-linear effects of the system, and a crude estimate of the
reduction is given by the formula
\begin{equation}
{p^*_{\rm eff}}=\frac{p^*}{u_0L}\int_L^{0_-}u\diff x.
\end{equation}

\subsection{Location of the spatial boundary}

In our approach, we assume turbulence is generated at the spatial boundary
by a non-resonant instability driven 
by particles escaping into the undisturbed upstream medium. 
The condition that these waves are strongly driven is
\begin{equation}
\label{NRcond}
\zeta M_{\rm A}^2 \geq 1,
\end{equation}
\citep{bell04}
where 
\begin{equation}
\label{jcrexpression}
 \zeta= \frac{j_{cr}p^*_{\rm eff}mc}{e\rho_0 u_0^2}
\end{equation}
and $M_{\rm A}$ is the Alfv\'en Mach number of the shock,
in the medium upstream of the boundary.

\begin{figure}
\centering
\vbox{
   \psfrag{py}[Bl][Bl][1.0][-0]{\hspace*{-0cm} {  $\zeta M^2$}}
   \psfrag{px}[Bl][Bl][1.0][-0]{\hspace*{-0cm} { $R$}}
\includegraphics[width=0.45\textwidth]{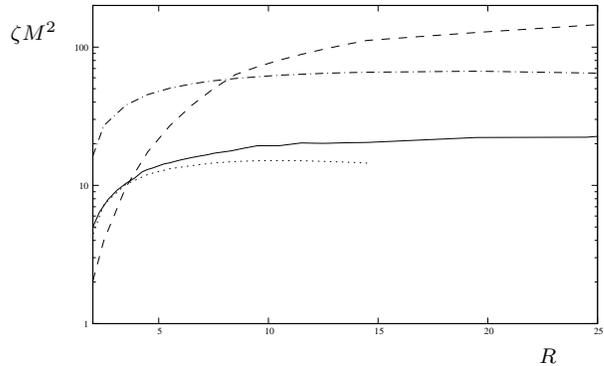}
}
\caption{ The driving parameter $\zeta$ times the square of the sonic
  Mach number $M$ as a function of subshock compression for different
  parameters.  Over the entire parameter range $\zeta
  M^2\gg1$. Therefore, provided the plasma $\beta\equiv M^2/M_{\rm
    A}^2$ is not too large, the non-resonant mode is, according to
  Eq.~(\ref{NRcond}), the fastest growing mode.  The lines correspond
  to $M=500$, $p^*=10^4$ (dashed), $M=100$, $p^*=10^5$ (solid),
  $M=100$, $p^*=10^4$ (dash-dot), $M=50$, $p^*=10^4$ (dotted), with an
  injection momentum of $p_0=0.1$.  }
\label{fig5}
\end{figure}

The cosmic-ray current is
evaluated from integration of the diffusive flux over momentum space
\begin{equation}
 j_{\rm cr}(x) = -4\pi e \int_{p_0}^\infty \kappa \pdiff{f}{x} p^2 \diff p.
\end{equation}
The integrand in this function peaks close to $p=p^*$ and can 
be easily extracted from the numerical solution to the 
full non-linear problem.
Because the magnetic field
ahead of the boundary is not specified in the solutions, we 
plot $\zeta M^2$ rather than $\zeta M_{\rm A}^2$
as a function of the precursor compression $R$ in 
Fig.~\ref{fig5}.
In the medium surrounding a 
supernova remnant, we expect $M_{\rm A}\sim M$. Therefore, according to this figure, 
the non-resonant waves are indeed strongly 
driven, as defined by Eq.~(\ref{NRcond}). In agreement with the linear theory,
we find that 
$j_{\rm cr}(-L)$
is an increasing function of the shock's Mach number
and a decreasing function of $p^*$.
For fixed maximum momentum, 
this result is independent of the the diffusion coefficient in the 
precursor, and, therefore, of the strength of the amplified magnetic field ---
a weaker field leads to a larger precursor but does not change the 
escaping particle flux. 

The assumption that particles escape freely upstream of the 
boundary implicitly assumes that the instability responsible for the
generation of the turbulence operates on a length scale that is short compared
to the precursor length. This requires that the inverse of the maximum growth rate 
$\gamma_{\rm max}$ of the non-resonant instability
should be less than the advection time through the precursor. 
The maximum growth rate is related to the driving parameter by
\eqb
\gamma_{\rm max}&=&{\zeta M_{\rm A}u_0\over 2 {r}_{\rm g}(p^*_{\rm eff})}
\eqe
where the gyro-radius of escaping particles is evaluated in 
the ambient magnetic field $B_0$ upstream of the boundary. 
It is convenient
to define the advection length
\eqb
L_{\rm adv} &=&  \frac{u_0}{\gamma_{\rm max}}.
\label{Ladv}
\eqe
However, the length scale in our numerical simulation is
set by the value of the diffusion coefficient in the amplified 
magnetic field $B_{\rm s}$. 
Relating this to the diffusion coefficient by assuming Bohm diffusion gives the following
lengthscale for our hydrodynamic precursor:
\eqb
L_{\rm pc}&=&{c {r}_{\rm g}(p^*)\over3 u_0}\left({B_0\over B_{\rm s}}\right).
\label{Lesc}
\eqe
The ratio of these two length scales is an important parameter 
\eqb
\label{ll1}
\frac{L_{\rm adv}}{L_{\rm pc}}&=&{6 \over \zeta} 
{v_{\rm A}\over c}\left({B_{\rm s}\over B_0}\right)
\left({p^*_{\rm eff}\over p^*}\right).
\eqe
and, in order to evaluate it, we must estimate the 
strength of the amplified field. 
\citet{bell04} and \citet{Pelletier}
give the following approximation for the saturated 
magnetic field energy density:
\eqb
\label{BS}
{B_{\rm s}^2\over 8\pi}&\approx& {3 u_1\over2c}P_{\rm CR}(0^-).
\eqe
Numerical investigations of the non-linear behavior of 
magnetic turbulence in the presence of streaming cosmic rays have been 
performed \citep[e.g.][]{Niemiec, Riquelme}, but the
results concerning the saturation level of this instability
are inconclusive. 
Adopting Eq.(\ref{BS}), and substituting into Eq.(\ref{ll1}) we find
\eqb
\frac{L_{\rm adv}}{L_{\rm pc}}&=&{6\sqrt{3}\over\zeta}
\left({u_0\over c}\right)^{3/2}
\left({P_{\rm CR}(0^-)\over R \rho_0 u_0^2}\right)^{1/2}
\left({p^*_{\rm eff}\over p^*}\right).
\label{advectcond}
\eqe

\begin{figure}
\centering
\vbox{
 \psfrag{py}[Bl][Bl][1.0][-0]{\hspace*{-0cm} { ${L_{\rm adv}\over L_{\rm pc}}$}}
 \psfrag{px}[Bl][Bl][1.0][-0]{\hspace*{-0cm} { $\nu$}}
\includegraphics[width=0.45\textwidth]{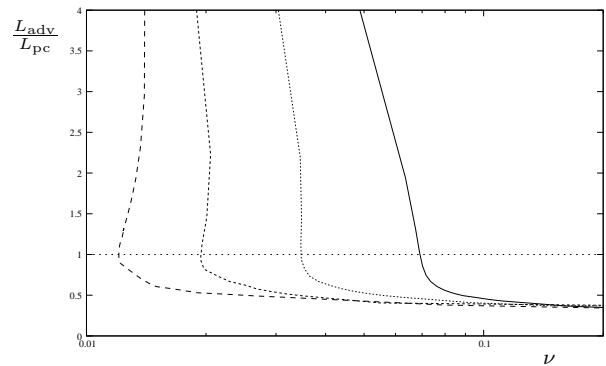}
}
\caption{%
The ratio of advection length to precursor length, $L_{\rm adv}/L_{\rm pc}$, 
for different values of $p^*$ for
a shock Mach number of $100$.
From left to right the lines correspond to $p^*=10^5$, $p^*=10^4$,
$p^*=10^3$, $p^*=10^2$. 
All models were calculated using an injection momentum of
$p_0=0.1$. The horizontal line corresponds to $L_{\rm adv} = L_{\rm pc}$.
}
\label{fig6}
\end{figure}

We plot $L_{\rm adv}/L_{\rm pc}$ determined numerically according to
Eq.(\ref{advectcond}) as a function of the injection parameter $\nu$
in Fig.~\ref{fig6}, for $M = 100$.  For weakly modified shocks with
$r_{\rm s}\approx 4$, the non-resonant instability grows slowly ahead
of the escape boundary: $L_{\rm adv}\gg L_{\rm pc}$.  For intermediate
strength modified shocks $4\lesssim R\lesssim 10$ (see Fig.~\ref{fig1}
to relate $\nu$ to $R$), the growth time is comparable to the
advection time.  For highly modified shocks, $R>10$, the instability
grows very rapidly in a small region just ahead of the escape
boundary.

Formally, the calculations we present are valid only in this latter
case, where the region in which the magnetic field increases from its
ambient strength to its strength in the precursor is small compared to
the length of the precursor itself. However, the point at which
amplification sets in is arbitrary in this case. The flux of particles
escaping from the shock front remains constant in planar geometry, so
that we should expect the pre-cursor length to increase as these
particles penetrate further and further upstream. This leads to a
higher $p^*$ and reduces the flux of escaping particles. In a fully
self-consistent picture, this process should regulate itself such that
the advection length becomes comparable to the precursor length.

On the other hand, for weakly modified shocks, where
$L_{\rm adv}\gg L_{\rm pc}$, the escaping particles
must penetrate a large distance in front of the shock before the
instability they drive has time to grow appreciably. In essence, this
is the situation investigated by \citet{ZirPtusk08}, who assumed the
precursor to the shock --- which is the region where the hydrodynamics
are influenced by the accelerated particles --- was very short
compared to the other length scales in the problem. According to our
results, this situation is consistent with stationarity of the
accelerated particle spectrum only for weakly modified shocks that are
relatively inefficient.  In this case, however, the magnetic field is
amplified very gradually, and the approximation that the amplification
region is short compared to the pre-cursor is inconsistent. This is
because a level of turbulence close to that assumed in the precursor
is already available and able to interact with particles far ahead of
the escape boundary. This again suggests that, in a self-consistent picture, 
the precursor length will be comparable to the advection
length of the instability.

According to Fig.~\ref{fig6}, the value of $\nu$ at which the ratio 
of $L_{\rm adv}$ to $L_{\rm pc}$ is unity
is a decreasing monotonic function of $p^*$, suggesting  
the physical picture evolves in the following manner: For a given injection
parameter $\nu$ and momentum $p^*$, determined from the length of the
precursor, we can calculate the ratio of the advection length to the
precursor length. If this ratio is large the escape boundary is too
far from the shock and $p^*$ will reduce itself until the ratio
reaches unity. If, however, the ratio is too small, $p^*$ will
increase, as described above.  In general, in the range of precursor
compression ratios that interest us here, $\nu$ is a decreasing
function of $p^*$ \citep{MalkovDrury}, which can be clearly seen in
Fig. \ref{fig2}.  It is natural to expect that the system will
organize itself such that the advection length approaches the
precursor length.  For the parameters adopted in Fig.~\ref{fig6},
this occurs for shocks in the intermediate range of modification.

This scenario suggests there exists
a relationship between the injection
and the maximum momentum in the system.
A similar connection has previously been investigated by \cite{MDV00}
in the context of self-organized criticality in cosmic-ray 
modified shocks,
although the value of the maximum momentum is controlled, in our
case, by a different mechanism. However, 
as in their case, the actual solution
depends on the microphysics at the subshock, which determines
the injection.

\section{Conclusions}
\label{conclusions}

The assumption underlying 
most previous investigations of non-linear diffusive shock acceleration
is the non-existence of waves  
able to scatter particles with energy above an upper cut-off. 
In this paper, 
we examine an alternative picture, in which the current
generated by the streaming cosmic rays falls below
some critical value at a distance $L_{\rm esc}$ upstream of the
subshock, and that beyond this the turbulence is insufficient to
scatter the particles.

In terms of the non-linear response of the system to changes in the 
injection parameter, we find the two pictures are quite similar
(see Fig.~\ref{fig1}). One difference is that our spatial boundary 
method can be used to model the shape of the distribution
close to the cut-off (see Fig.~\ref{fig3} and Fig.~\ref{fig4}). This is an advantage 
because it potentially enables 
one to model the radiative signatures of the acceleration process. 
 
The main difference, however, is that we are able to address the physics 
that determines the location of the boundary. Previous work that implemented
such a boundary \cite{VEB06} did not constrain its location.
In time-dependent models of 
acceleration in supernova remnants \cite{berezhkoetal97}, 
an effective spatial boundary is imposed by the spherical geometry.
But they assume a value for the amplified magnetic field in the 
entire computational box,
without considering whether or not this is consistent with the location of the 
boundary. 

We argue that the location of the boundary 
is determined by the growth rate of the 
instability responsible for field amplification.
It has recently been shown that, in the case of
efficient shock acceleration, 
a short-wavelength non-resonant mode \cite{bell04} plays a crucial role.
In a series of papers,
\cite{Pelletier,Marcowith} argue that this non-resonant
mode dominates far from the shock, while the resonant
streaming instability takes over closer to the shock, driving 
the diffusion towards Bohm-type in the precursor. 
According to these arguments, the position of the free-escape boundary 
should be fixed by the properties of the non-resonant mode, once 
the shock is modified by the cosmic-ray pressure. 
We show explicitly that the non-resonant modes are strongly driven 
at the escape boundary, and compare the local growth rate with the 
rate at which the waves are advected towards the shock front 
(see Fig.~\ref{fig6}). 
In this picture, particles 
stream freely ahead (upstream) of the boundary, 
whereas behind (downstream of) it, the 
turbulence is assumed to be 
fully developed and particles undergo Bohm diffusion 
in an amplified magnetic field.

\cite{ZirPtusk08} also suggest that
the geometry of a supernova shock front sets the length scale for the 
precursor, and thus determines the maximum particle energy. 
However, we find that when the non-linear dynamics of the acceleration process
are included, the length scale is set by the strength of the amplified 
magnetic field and the efficiency of the injection process.
Ultimately, it is the microphysics of injection --- represented 
in Figs.~\ref{fig1} and \ref{fig6} by the parameter 
$\nu$ --- 
that determines not only the efficiency of the acceleration process, but
also the maximum attainable particle energy.

\acknowledgments
This research was supported by a Grant from the G.I.F., the German-Israeli
Foundation for Scientific Research and Development. PD thanks the MPI for Nuclear 
Physics for their hospitality during this work. BR gratefully acknowledges 
support from the Alexander von Humboldt foundation.

\end{document}